\documentclass[%
reprint,
superscriptaddress,
amsmath,amssymb,
aps,pre,
floatfix,
]{revtex4-1}

\usepackage{graphicx}
\usepackage{dcolumn}
\usepackage{bm}
\usepackage{mathtools}

\newcommand{\HEIGHT}{\Delta E_{\text{r}}}

\begin{document}
	
	\preprint{APS/123-QED}
	
	\title{Breaking time-reversal symmetry for ratchet models of molecular machines}
	
	\author{Arshia Zarrin}
	\affiliation{Department of Physics, Simon Fraser University, Burnaby, British Columbia, Canada}
	\author{David A.~Sivak}
	\email{dsivak@sfu.ca}
	\affiliation{Department of Physics, Simon Fraser University, Burnaby, British Columbia, Canada}
	\author{Aidan I.~Brown}
	\email{aibrown@ucsd.edu}
	\affiliation{Department of Physics, Simon Fraser University, Burnaby, British Columbia, Canada}
	\affiliation{Department of Physics, University of California, San Diego, San Diego, California, USA}
	\date{\today}
	
	\begin{abstract}
		Biomolecular machines transduce free energy from one form to another to fulfill many important roles inside cells, with dissipation required to achieve directed progress. We investigate how to break time-reversal symmetry at a given dissipation cost by using deterministic protocols to drive systems over sawtooth potentials, which have frequently been used to model molecular machines as ratchets. Time asymmetry increases for sawtooth potentials with higher barriers and for driving potentials of intermediate width. For systems driven over a sawtooth potential according to a protocol, we find that symmetric sawtooths maximize time asymmetry, while earlier work examining ratchet models of molecular machines required asymmetric sawtooth potentials to achieve directed behavior. This distinction arises because deterministically driven machines are externally provided with direction, while autonomous machines must generate directed behavior.
	\end{abstract}
	
	\maketitle

	\section{Introduction}
	
	Biomolecular machines perform a variety of essential tasks inside living cells~\cite{alberts98}, and inspire the development of synthetic molecular machines~\cite{zhang18}. Molecular machines operate out of equilibrium and stochastically~\cite{seifert12}, buffeted by fluctuations at the nanoscale~\cite{brown17b}. As with macroscopic machines, molecular machines can be designed to optimize efficiency~\cite{vandenbroeck12,schmiedl08} and speed~\cite{wagoner16,brown17,brown18}. However, unlike their macroscopic counterparts, design can meaningfully impact the precision of molecular machine operation~\cite{barato15}.
	
	Molecular machines can pay a significant fraction of their free energy budget for directed progress~\cite{machta15,ariga18}. To quantify directional behavior, Feng and Crooks~\cite{feng08} proposed time-reversal asymmetry, expressed as the Jensen-Shannon divergence between trajectory distributions for processes driven in forward and reverse directions.  
	
	Earlier investigations (unfolding/refolding of ribonucleic acid hairpins in experiment~\cite{feng08} and alanine decapeptide in molecular dynamics simulations~\cite{procacci10}) found the trade-off between time asymmetry (seen as a potential benefit) and dissipation (a cost) remained near the prediction of linear-response theory~\cite{marconi08}. Linear-response behavior for time asymmetry and dissipation is expected for slow processes that remain near equilibrium.
	For a range of dissipation (approximately 1-8 $k_{\text{B}}T$), there is a substantial gap between the linear-response prediction for time asymmetry and the maximum possible time asymmetry, while for lower and higher dissipation the difference between linear-response and the maximum possible time asymmetry is small. This gap (at medium dissipation) presents an opportunity for significant improvement on linear-response behavior~\cite{brown16}. We aim to determine characteristics that allow time asymmetry to increase for a given dissipation cost, or equivalently decrease dissipation for a given time asymmetry.
	
	We investigate the trade-off between time asymmetry and dissipation in the context of molecular machines, using a sawtooth potential that is frequently used to represent a ratchet model of a molecular machine~\cite{astumian94,julicher97,astumian98,parmeggiani99,lipowsky00,reimann02,suzuki03,gabrys04,linke05,lacoste09,levien15,kedem17,saakian18}. `Flashing' ratchet models, which switch between multiple potential energy landscapes, have been used to generate autonomous directed motion~\cite{suzuki03}. While we use the sawtooth landscape from flashing ratchets, we apply an externally directed protocol (in contrast to autonomous behavior) to produce directed motion. A protocol typically represents the influence of an experimental apparatus, providing a suitable description of externally driven synthetic molecular machines, with more limited direct application to autonomous biomolecular machines.
	Understanding how molecular machines can achieve higher time asymmetry for a given dissipation cost may point towards design principles for molecular machines. These principles could inform the design of more efficient synthetic molecular machines.
	
	In our previous work~\cite{brown16}, we explored how to manipulate time asymmetry for a step potential representing energy transmission, using a harmonic trap to move the system from one side of the step to the other. Intermediate step heights and protocols remaining near the step led to time asymmetries that improved most on linear response.
	
	In contrast to step potentials, we find that high sawtooth potentials provide greater improvement over linear response. Changing from a symmetric to an asymmetric sawtooth causes time asymmetry to decrease, stemming from the greater overlap in initial equilibrium distributions for more asymmetric sawtooths.

	\section{Methods}
	Time-reversal asymmetry $A$ (here measured in nats, $1/\ln 2$ bits) quantifies the expected information gain about whether a single observed trajectory was produced by a particular nonequilibrium process or its time reversal, assuming a 50\% prior probability of the forward or reverse direction~\cite{crooks11}. If the forward and reverse trajectory distributions are identical, then any observation provides no information on the protocol direction, defining a minimum $A = 0$. If the forward and reverse trajectory distributions are completely distinct, then any observed trajectory can be definitively assigned to the forward or reverse process, defining a maximum $A = \ln 2$ nats, amounting to one bit of information.
	
	The Crooks fluctuation theorem~\cite{crooks99} quantitatively describes the relative probabilities of forward and reverse trajectories solely in terms of the work required for each trajectory. This permits expression of the time asymmetry $A$ in terms of the forward and reverse work distributions~\cite{feng08},
	\begin{align}
	\label{eq:asymmetry}
	A &= \frac{1}{2}\left\langle\ln\frac{2}{1 + \exp\left(-\beta W[X|\Lambda] + \beta\Delta F\right)}\right\rangle_{\Lambda}\nonumber\\
	&+\frac{1}{2}\left\langle\ln\frac{2}{1 + \exp(-\beta W[\tilde{X}|\tilde{\Lambda}] - \beta\Delta F)}\right\rangle_{\tilde{\Lambda}} \ .
	\end{align}
	$\Lambda$ labels a protocol, the time schedule of a controllable parameter $\lambda(t)$ over $t\in[0,\Delta t]$ for a protocol of duration $\Delta t$. The reverse protocol $\tilde{\Lambda}$ is the time-reversal of $\Lambda$, i.e.\ $\lambda(\Delta t-t)$. $W[X|\Lambda]$ is the work done during protocol $\Lambda$ when the system follows trajectory $X$. 
	$\Delta F$ is the free energy change during protocol $\Lambda$, and $\beta \equiv (k_{\text{B}}T)^{-1}$, for Boltzmann's constant $k_{\text{B}}$ and temperature $T$.
	
	Dissipation $h$ is the average excess work for an equal weighting of forward and reverse protocols, and hence equals the average work because the equilibrium free energy changes during $\Lambda$ and $\tilde{\Lambda}$ cancel~\cite{feng08}:
	\begin{equation}
	\label{eq:dissipation}
	h[\Lambda] \equiv \frac{1}{2}\left(\beta\langle W[X|\Lambda]\rangle_{\Lambda} + \beta\langle W[\tilde{X}|\tilde{\Lambda}]\rangle_{\tilde{\Lambda}}\right) \ .
	\end{equation}
	
	Time asymmetry is mathematically upper bounded by $A\leq h/4$ and $A\leq \ln[2/(1+e^{-h})]$~\cite{feng08,taneja05}. A system which remains near equilibrium responds linearly to small perturbations~\cite{chandler87} -- for a driven ensemble in this linear-response regime, the work distribution is Gaussian, and the Crooks fluctuation theorem imposes a connection between the mean and variance of the work~\cite{feng08,speck04}, $\langle W\rangle - \Delta F = \frac{1}{2}\sigma_W^2$.
	
\begin{figure}[tbp] 
		\centering
		\hspace{-0.0in}
		\begin{tabular}{c}        
			\hspace{-0.150in}
			\includegraphics[width=3.50in]{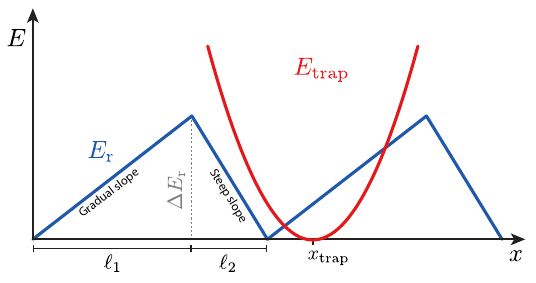}
		\end{tabular}
		\caption{\label{fig:potential} 
			{\bf Sawtooth energy landscape model}. The potential landscape is composed of a time-dependent quadratic trap, $E_{\text{trap}}[x,x_{\text{trap}}(t)] = \tfrac{1}{2}k[x-x_{\text{trap}}(t)]^2$ (red); and a time-independent sawtooth potential $E_{\text{r}}(x)$ (blue) from Eq.~\eqref{eq:ratchetpotential}. When the sawtooth is asymmetric, the gradual ($\ell_i > 0.5$) and steep ($\ell_i < 0.5$) sides of the sawtooth are indicated.} 
	\end{figure}

	We investigate time asymmetry and dissipation with an overdamped system of diffusivity $D$ on a potential landscape $E[x,x_{\text{trap}}(t)] = E_{\text{trap}}[x,x_{\text{trap}}(t)] + E_{\text{r}}(x)$. The first component is a quadratic trap, $E_{\text{trap}}[x,x_{\text{trap}}(t)] = \frac{1}{2}k[x - x_{\text{trap}}(t)]^2$, with a time-dependent minimum $\lambda(t) = x_{\text{trap}}(t)$ as the control parameter. The second component $E_{\text{r}}(x)$ is a time-independent periodic sawtooth (ratchet) potential (Fig.~\ref{fig:potential}) that enables directional motion for simple models of molecular machines~\cite{astumian94,julicher97,astumian98,parmeggiani99,lipowsky00,reimann02,suzuki03,gabrys04,linke05,lacoste09,levien15,kedem17,saakian18},
	\begin{equation}
	\label{eq:ratchetpotential}
	E_{\text{r}}(x) =
	\begin{cases}
	\HEIGHT\frac{x\text{ mod }\ell}{\ell_1} \ , &\text{if } 0 < x \text{ mod }\ell\leq\ell_1\\
	\HEIGHT \frac{\ell - x\text{ mod }\ell}{\ell_2} \ , &\text{if } \ell_1 < x \text{ mod } \ell \leq \ell
	\end{cases}
	\ ,
	\end{equation}
	with $\ell = \ell_1 + \ell_2$ the wavelength over which the sawtooth repeats, and $\HEIGHT$ the barrier height. The periodicity of $E_{\text{r}}(x)$ represents a cyclical molecular machine, with $\Delta F=0$ over a cycle. $E$, $k$, $x$, $\HEIGHT$, and $\Delta t$ are nondimensionalized by energy scale $k_{\text{B}}T$, length scale $\ell$, and a timescale such that system diffusivity $D=1$ (see Appendix~\ref{sec:nondimensionalization} for details).
	
	For each protocol, the trap translates at a constant velocity $u=1/\Delta t$ from $x_{\text{trap}}=0$ to $x_{\text{trap}}=1$ for the forward protocol, and in the opposite direction for the reverse protocol. The system begins in equilibrium and diffuses as the trap translates. Appendix~\ref{sec:simulation} provides further simulation details.
	
	For a given spring constant $k$, ratchet height $\HEIGHT$, sawtooth lengths $\ell_1$ and $\ell_2$, and trap speed $u$, work distributions are compiled from many repetitions of forward and reverse protocols, from which a single time asymmetry [Eq.~\eqref{eq:asymmetry}] and dissipation [Eq.~\eqref{eq:dissipation}] are calculated.

	\section{Results}
	
	\begin{figure*}[tbp]
		\centering
		\hspace{-0.0in}
		\begin{tabular}{c}
			\hspace{-0.2in}\includegraphics[width=7.3in]{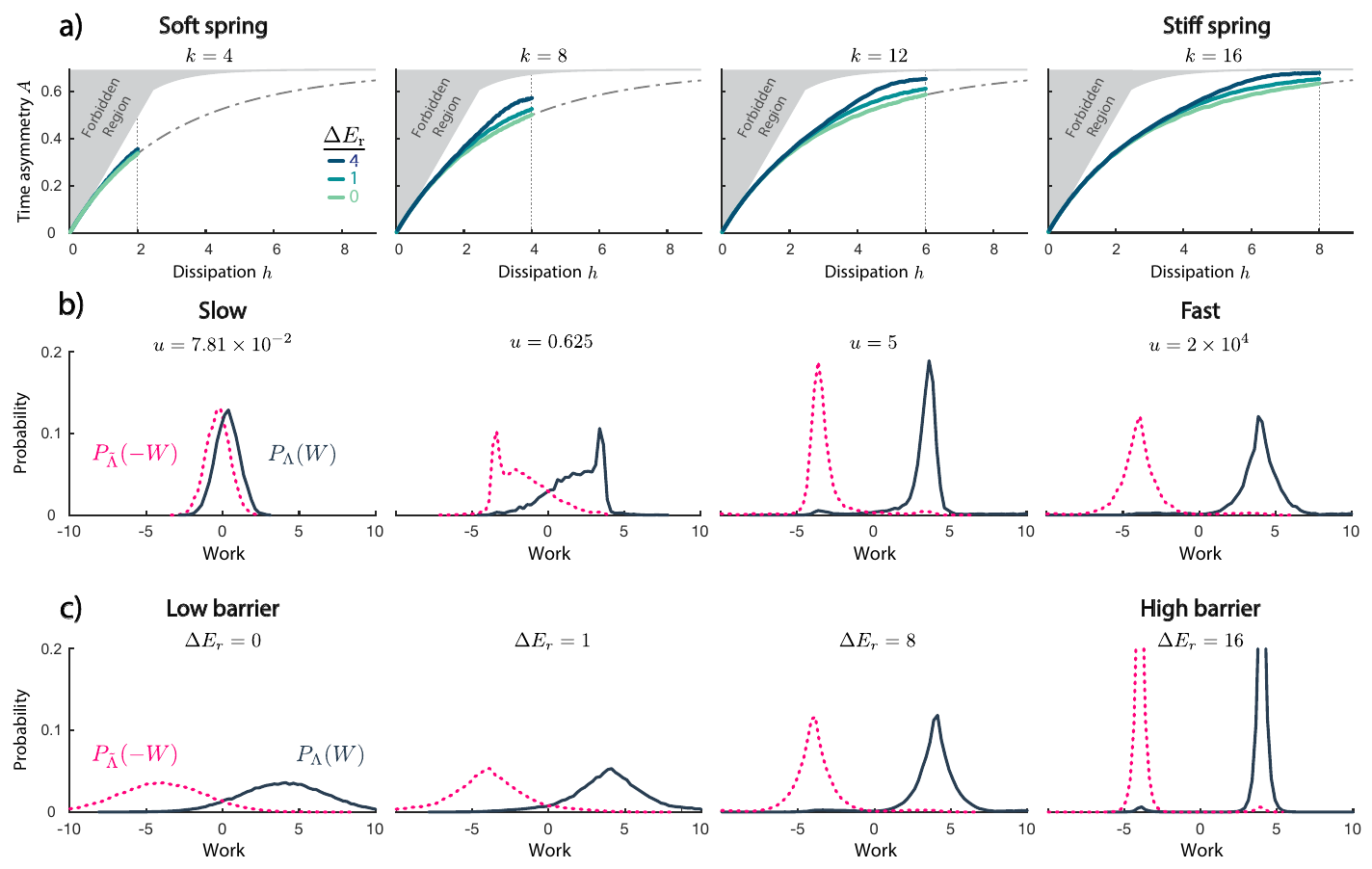}
		\end{tabular}
		\caption{\label{fig:Avsh} 
			{\bf Time asymmetry and work distributions} for symmetric ratchet potentials ($\ell_1=\ell_2 = 0.5$).
			a) Time asymmetry $A$ vs.\ dissipation $h$, with each curve generated by parametrically varying trap speed. Trap spring constant $k$ increases from left panel to right panel. Curve color indicates barrier height $\HEIGHT$. Gray region indicates forbidden time asymmetries~\cite{feng08,taneja05}. Dashed-dotted curve indicates linear-response behavior. Vertical dashed line indicates maximum dissipation $h_{\text{max}}$ for a given spring constant.
			b,c) Work distributions for forward (black solid curves) and reverse (red dashed) protocols. 
			b) Trap speed $u$ increases from left panel to right. Spring constant $k=8$ and barrier height $\HEIGHT = 4$.
			c) Barrier height $\HEIGHT$ increases from left panel to right. Spring constant $k=8$ and trap speed $u=2\times10^4$. Points in a) and distributions in b) are over $5\times10^4$ samples. Distributions in c) are over $10^5$ samples.
		}  
	\end{figure*}

	\subsection{Symmetric ratchets}
	We first study symmetric sawtooth potentials, $\ell_1 = \ell_2 = 0.5$. Fig.~\ref{fig:Avsh}a displays curves of time asymmetry vs.\ dissipation, with each curve generated by parametrically varying the trap speed $u$. As trap speed $u$ increases, both the time asymmetry and the dissipation increase in a manner that depends on the barrier height $\HEIGHT$ and spring constant $k$. For each curve, both the time asymmetry and dissipation are negligible for low trap speed $u$ (bottom left of Fig.~\ref{fig:Avsh}a panels). A slow-moving trap allows the system to stay near equilibrium, leading to forward and reverse work distributions with similar shapes, widths, and mean work magnitudes (Fig.~\ref{fig:Avsh}b, far left panel). These near-equilibrium trajectories are well-described by linear response, hence staying near the linear-response time asymmetry-dissipation curve. 
	
	As $u$ increases (left to right in Fig.~\ref{fig:Avsh}b), the forward and reverse work distributions become non-Gaussian and more separated, leading to less overlap and causing both time asymmetry and dissipation to increase. For these higher trap speeds, the system departs from linear response (highest dissipations in each panel of Fig.~\ref{fig:Avsh}a).
	
	At higher trap speeds, individual system trajectories can be roughly categorized as either crossing from the initial side to the far side of the sawtooth potential barrier, or remaining on the initial side without crossing the barrier. An activated process~\cite{grabert88} occurs at a typical rate $k_{\text{a}} = \tfrac{2D}{(\Delta x^{\ddagger})^2}e^{-\Delta E^{\ddagger}}$ for respective distance $\Delta x^{\ddagger}=\ell/2$ and energy difference $\Delta E^{\ddagger}$ from the minimum to the barrier. 
	For $k\leq 2\HEIGHT/\ell^2$ the energy minimum on the initial side of the barrier remains at $x=0$ and the energy barrier remains at $x=\ell/2$ regardless of $x_{\text{trap}}$, but the barrier height varies as the trap moves, 
	giving 
	\begin{subequations}
		\begin{align}
		\Delta E^{\ddagger}(x_{\text{trap}}) &\equiv E^{\ddagger}(x_{\text{trap}}) - E_{\text{min}}(x_{\text{trap}}) \\
		&= \left[\HEIGHT + \tfrac{1}{2}k\left(x_{\text{trap}} - \tfrac{\ell}{2}\right)^2 \right] - \tfrac{1}{2}kx_{\text{trap}}^2 \\
		&= \HEIGHT -\tfrac{1}{2}k\ell\left(x_{\text{trap}}-\tfrac{\ell}{4}\right) \ .
		\end{align}
	\end{subequations}
	For average crossing rate $\langle k_{\text{cross}}\rangle = \tfrac{1}{\ell} \, \int_0^{\ell} k_{\text{a}}(x_{\text{trap}}) \, \text{d}x_{\text{trap}}$ this gives the barrier crossing timescale for the protocol as a whole:
	\begin{subequations}
		\begin{align}
		t_{\text{cross}} &\equiv \langle k_{\text{cross}}\rangle^{-1} \\ 
		&= \frac{k\ell^4}{16D} \frac{e^{\HEIGHT+k\ell^2/8}}{e^{k\ell^2/2} - 1} \ .
		\end{align}
	\end{subequations}
	The leftmost panel of Fig.~\ref{fig:Avsh}b ($u=7.81\times10^{-2}$) corresponds to $t_{\text{cross}}/\Delta t \approx 0.11$, allowing the system to reliably cross the barrier during the protocol duration, leading to a single component in the work distributions.
	
	Faster driving ($u=0.625$ in Fig.~\ref{fig:Avsh}b) corresponds to $t_{\text{cross}}/\Delta t \approx 0.86$.
	At this driving speed the system does not reliably cross the barrier during the protocol duration, and the work distributions contain two distinct components, corresponding to 
	two trajectory categories. The wide `crossing' components in Fig.~\ref{fig:Avsh}, at low work magnitude, represent systems that cross to the far side of the barrier, thus staying closer to $x_{\text{trap}}$, such that the trap does less work on the system. The narrow `non-crossing' peaks, at higher work magnitude, are for systems that do not have sufficient time to cross to the far side of the barrier, keeping the system further from $x_{\text{trap}}$ and increasing the work done by the trap on the system.
	
	As the trap speed $u$ increases further, the system is more likely to remain on the initial side of the barrier without crossing, shifting probability from the crossing component to the non-crossing component (Fig.~\ref{fig:Avsh}b, $u=5$ and $u=2\times10^4$).
	At the highest trap speeds, the protocol duration is insufficient for substantial system diffusion, so the system never crosses to the far side of the barrier.
	
	As trap speed $u$ increases (Fig.~\ref{fig:Avsh}b, left to right), work distributions depart from Gaussian distributions, gaining the secondary components discussed in the preceding paragraphs. These components, with less overlap than the Gaussian work distributions of linear response, cause time asymmetries above linear response, as shown in Fig.~\ref{fig:Avsh}a.
	
	For higher barrier $\HEIGHT$, time asymmetry $A$ is higher at a given dissipation (Fig.~\ref{fig:Avsh}a). This can be considered as either increasing the time asymmetry at a given dissipation, or reaching a given time asymmetry at a lower dissipation. Figure~\ref{fig:Avsh}c shows work distributions for increasing barrier height $\HEIGHT$. As $\HEIGHT$ increases, the system is less likely to cross from the initial to the far side of the barrier, which narrows the work distributions and decreases their overlap, producing higher time asymmetries. 
	
	Maximum dissipation is achieved with the fastest trap speeds, approaching instantaneous trap movement, such that the system is unable to substantially diffuse over the protocol duration. The maximum achievable time asymmetry increases with trap stiffness $k$. For a system effectively frozen in its initial position (drawn from the initial equilibrium ensemble), the maximum dissipation $h_{\text{max}}=\tfrac{1}{2}k$ (shown as a vertical dashed line in Fig.~\ref{fig:Avsh}a) equals the energy change in changing the trap from $x_{\text{trap}}=0$ to $x_{\text{trap}}=1$, averaged over the equilibrium distribution given $x_{\text{trap}}=0$ (see Appendix~\ref{sec:instant}).
	
	\begin{figure}[tbp] 
		\centering
		\hspace{-0.0in}
		\begin{tabular}{c}
			\hspace{-0.150in}\includegraphics[width=\columnwidth]{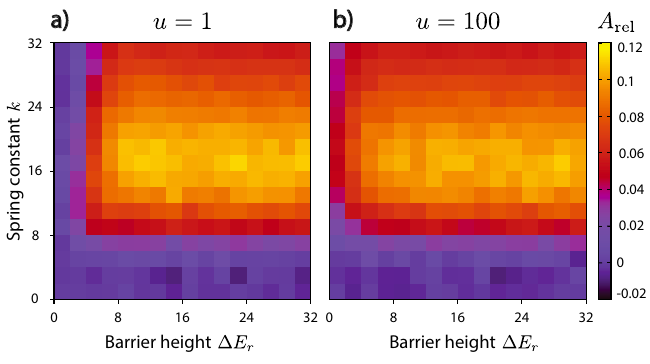}
		\end{tabular}
		\caption{\label{fig:colormaps} 
			{\bf Maximizing time asymmetry} above linear response for symmetric potentials. Color map indicates $A_{\text{rel}}(h) \equiv A(h) - A_{\text{LR}}(h)$, the time asymmetry above that of linear response at the corresponding dissipation, calculated from $10^4$ samples. 
			a) Relatively slow trap speed $u=1$.
			b) Faster trap speed $u=100$. 
		}  
	\end{figure}
	
	As linear-response behavior is achieved by relatively slow processes that remain near equilibrium and in earlier empirical investigations of time asymmetry across a range of speeds and processes~\cite{feng08,procacci10}, we seek driving characteristics that exceed linear-response time asymmetries for a given dissipation. Figure~\ref{fig:colormaps} shows time asymmetry $A$ relative to linear-response time asymmetry $A_{\text{LR}}$ at the same dissipation, $A_{\text{rel}}(h) \equiv A(h) - A_{\text{LR}}(h)$, for two speeds $u$ as spring constant $k$ and barrier height $\HEIGHT$ are varied. $A_{\text{rel}}$ quantifies the improvement of time asymmetry upon the linear response time asymmetry at a given dissipation, distinct from the absolute time asymmetry $A$.
	
	The dependence of $A_{\text{rel}}$ on $k$ is unchanged across a wide range of $u$, with $A_{\text{rel}} \approx 0$ for low $k$ and $A_{\text{rel}}$ sharply increasing to substantial positive values at $k\approx10$, provided $\HEIGHT$ is not too small. The characteristic width $\delta x$ of system fluctuations in the trap scales as $\delta x = k^{-1/2}$, and at $k\approx10$ is $\delta x/\ell
	\approx 0.3$.
	
	$A_{\text{rel}} \approx 0$ for low $\HEIGHT$ and sharply increases at a threshold value of $\HEIGHT$, provided $k \gtrapprox 10$. The threshold $\HEIGHT$ value for the increase in $A_{\text{rel}}$ decreases as $u$ increases. For $u=1$, 
	$A_{\text{rel}}$ increases at $\HEIGHT\approx 4$, corresponding to $t_{\text{cross}}/\Delta t \approx 1.4$.
	This barrier-crossing timescale is slightly longer than the protocol duration and is similar to relative speeds necessary for substantial non-crossing trajectories in Fig.~\ref{fig:Avsh}. For $u=100$ (protocol duration of 0.01), the $\HEIGHT$ necessary to increase $A_{\text{rel}}$ drops to nearly zero. Even the barrier-crossing timescale at $\HEIGHT=1$ is much longer than this protocol duration.
	
	In Fig.~\ref{fig:colormaps}, $A_{\text{rel}} > 0$ for traps significantly tighter than the sawtooth wavelength and for barrier crossing timescales much longer than the protocol duration. These conditions combine to limit barrier crossing (which implies more distinction between forward and reverse trajectory distributions) and increase dissipation, allowing $A_{\text{rel}}$ to rise.
	
	For larger $\HEIGHT$, there is an optimum $k\approx18$ which maximizes $A_{\text{rel}}$ (Fig.~\ref{fig:colormaps}). At low and high dissipation, linear-response time asymmetry nears the maximum possible time asymmetry, limiting $A_{\text{rel}}$. At intermediate dissipation, there is substantial room between linear response and forbidden time asymmetries, permitting larger $A_{\text{rel}}$. Intermediate $k$ values select these intermediate dissipation values that permit large $A_{\text{rel}}$.
	
	After an initial sharp increase of $A_{\text{rel}}$ at a threshold $\HEIGHT$, $A_{\text{rel}}$ plateaus with further increases of $\HEIGHT$. This plateau underlies the diminishing returns in increasing time asymmetry for increasing $\HEIGHT$ in Fig.~\ref{fig:Avsh}a, caused once $\HEIGHT$ is sufficiently large to prevent nearly all system trajectories from crossing the barrier.

	\subsection{Asymmetric ratchets}
\begin{figure}[!t] 
		\centering
		\hspace{-0.0in}
		\begin{tabular}{c}
			\hspace{-0.150in}\includegraphics[width=\columnwidth]{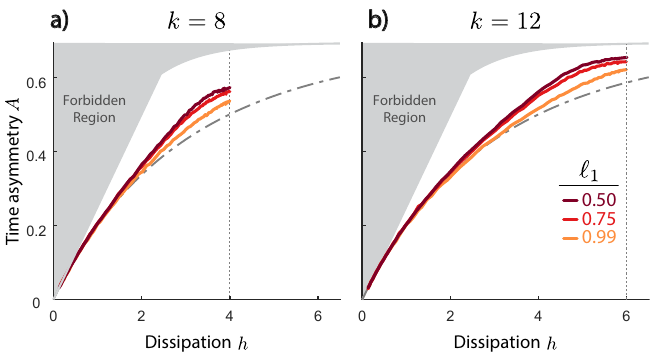}
		\end{tabular}
		\caption{\label{fig:asymmetry} 
			{\bf Time asymmetry for asymmetric ratchets.} Time asymmetry $A$ and dissipation $h$ for asymmetric sawtooth potentials, with different curves corresponding to different $\ell_1 \geq 0.5$.
			($\HEIGHT=4$.) a) Spring constant $k=8$. b) $k=12$. Gray region shows time asymmetries forbidden for a given dissipation. Dashed-dotted curve shows linear-response behavior. Vertical dotted lines indicate $h_{\text{max}} = \tfrac{1}{2}k$. Each point is calculated from $5\times10^4$ samples.
		}  
	\end{figure}

	We now explore asymmetric ratchets with $\ell_1 \neq \ell_2$. Figure~\ref{fig:asymmetry} shows time asymmetry vs.\ dissipation for various sawtooth asymmetries. Potentials with $\ell_1=0.5 \pm \delta$, for a given $\delta \in [0,0.5]$, will yield the same time asymmetry, because time asymmetry does not depend on which work distribution is labeled forward and which is labeled reverse. Accordingly, we increase $\ell_1$ from 0.5 towards 1 to explore the effect of an asymmetric potential. We find that time asymmetry $A$ at a given dissipation decreases as the asymmetry of the potential increases.
	
\begin{figure*}[t]
		\centering
		\hspace{-0.0in}
		\begin{tabular}{c}
			\hspace{-0.2in}\includegraphics[width=7.3in]{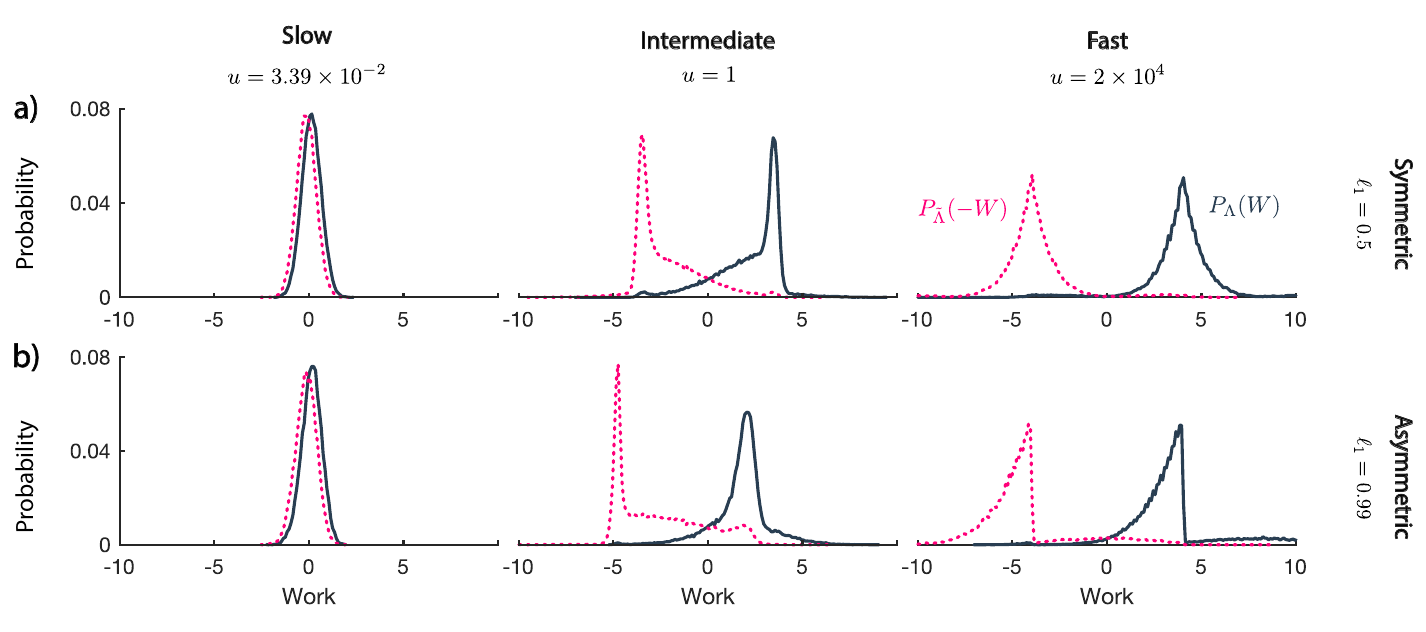}
		\end{tabular}
		\caption{\label{fig:AsymmetryWorkDists} 
			{\bf Work distributions} for forward (black) and reverse (red) protocols with 
			a) symmetric sawtooth potentials ($\ell_1 = \ell_2 = 0.5$), or
			b) asymmetric sawtooth potentials ($\ell_1 = 0.99$ and $\ell_2 = 0.01$). Driving speed increases from left to right panels. Spring constant $k=8$ and barrier height $\HEIGHT=4$. Distributions are over $5\times10^4$ samples.
		}  
	\end{figure*}

	To understand the time asymmetry decrease as $\ell_1$ increases, Fig.~\ref{fig:AsymmetryWorkDists} compares work distributions for symmetric ($\ell_1=\ell_2$) and asymmetric ($\ell_1 \neq \ell_2$) potentials. For slow trap speed, the work distributions for symmetric and asymmetric ratchets are similar and appear Gaussian (Fig.~\ref{fig:AsymmetryWorkDists}, far left). As trap speed increases, the work distributions for symmetric and asymmetric ratchets become distinct.
	
	A symmetric sawtooth potential and an intermediate trap speed (Fig.~\ref{fig:AsymmetryWorkDists}a, middle) lead to non-Gaussian forward and reverse work distributions, with a clear peak corresponding to trajectories that do not cross the barrier, and a tail of trajectories that cross the barrier. For a comparable asymmetric ratchet, the forward and reverse work distributions are not symmetric (Fig.~\ref{fig:AsymmetryWorkDists}b, middle). The forward-protocol peak is wider, as $\ell_1 > 0.5$ allows the system during the forward protocol to slowly climb up the gradual potential slope (Fig.~\ref{fig:potential}). The primary reverse-protocol peak is narrower, as the system is unable to make the steep climb; a secondary component of the reverse-protocol work distribution occurs at the same work value as the forward-protocol peak, due to trajectories which stay on the gradually sloped side of the sawtooth. The overlap of the forward-protocol peak and reverse-protocol secondary component decrease the time asymmetry compared to the symmetric work distributions. Compared to symmetric potentials, asymmetric sawtooth potentials increase the probability that the system is on the gradual side of the ratchet potential (independent of whether the trap starts on the steep or gradual side), leading to more similar trajectories overall, and thus more difficulty in distinguishing between the forward and reverse protocols.
	
	Fast driving of a symmetric ratchet produces forward and reverse work distributions composed of single, symmetric, non-Gaussian peaks (Fig.~\ref{fig:AsymmetryWorkDists}a, right). For asymmetric potentials, the work distribution peaks nearly recapitulate the sawtooth potential shape, with one side rising gradually and the other quite steeply (Fig.~\ref{fig:AsymmetryWorkDists}b, right). The work distributions for both symmetric and asymmetric potentials are caused by the initial equilibrium distributions, as there is insufficient time for substantial system diffusion. In addition to the primary peak in each distribution, there is a much smaller secondary component at more positive work values -- the reverse-protocol secondary component overlaps with the forward-protocol primary peak. As for intermediate speeds, the greater overlap of the work distributions for the asymmetric sawtooth leads to a lower time asymmetry. The cause of this greater overlap in work distributions is a greater overlap in initial equilibrium positions: a system with a trap which begins the protocol centered on the steep side of the potential (reverse protocol) has a non-vanishing equilibrium probability to be on the gradual side of the potential.
	
	Overall, in an asymmetric ratchet the system is more likely to be in the same initial equilibrium position during forward and reverse protocols, compared to symmetric ratchets. Slower protocols that allow time for diffusion eliminate this difference in overlap of initial positions reflected in the work distributions.

	\section{Discussion}
	We have investigated the trade-offs of time asymmetry and dissipation for driven systems diffusing in sawtooth potentials, which are frequently used as simple ratchet models for molecular machines~\cite{astumian94,julicher97,astumian98,parmeggiani99,lipowsky00,reimann02,suzuki03,gabrys04,linke05,lacoste09,levien15,kedem17,saakian18}. Accordingly, we explored the effect of sawtooth potential (height and asymmetry) and protocol (speed and driving strength) characteristics on time asymmetry. We have described how these factors lead to a relatively high time asymmetry for a given dissipation, or equivalently how to reduce the dissipation necessary to achieve a given time asymmetry.
	
	For high barriers and intermediate spring constants, high trap velocities lead to time asymmetries well above linear response for a sawtooth potential (Fig.~\ref{fig:Avsh}a). The barrier is difficult to cross in either direction (as the system must climb the barrier), such that for fast protocols the system rarely crosses the barrier during the entire protocol. This finding suggests molecular machines could gain a higher time asymmetry by proceeding through much of each cycle rapidly, consistent with recent work describing the kinesin cycle with slow transitions following rapid events, such as ATP-induced isomerization of kinesin following ATP binding~\cite{chang13,isojima16}, diffusive search for a forward microtubule binding site following neck linker docking~\cite{isojima16}, and ADP release following binding of the kinesin head to the microtubule~\cite{hackney88,nogales16}.
	
	Increasing the barrier height of a sawtooth potential increases time asymmetry for a given dissipation, until the barrier is sufficiently high that the system rarely crosses the barrier (Figs.~\ref{fig:Avsh} and \ref{fig:colormaps}). The probability of crossing the barrier decreases as the barrier height increases, leading to more distinct forward and reverse trajectory distributions.
	
	The linear-response relationship between time asymmetry and dissipation holds for slow processes that allow the system to stay near equilibrium~\cite{speck04}. To assess performance for driving the system over a sawtooth potential, we compared time asymmetries for the sawtooth to time asymmetries for linear-response behavior at the same dissipation (Fig.~\ref{fig:colormaps}), finding a clear optimum spring constant which maximizes the improvement over linear response. At this optimum spring constant, relatively fast protocols produce dissipation in the range ($1 k_{\text{B}}T\lesssim h\lesssim 8$~$k_{\text{B}}T$) at which linear-response time asymmetry is far below forbidden time asymmetries. This optimum spring constant suggests that molecular machines may benefit from states that allow some meaningful position fluctuations without high energetic cost, as long as these fluctuations are not so large as to be comparable to the separation between states. These states --- that are not too tightly or loosely constrained --- are consistent with the somewhat constrained distributions of kinesin position~\cite{isojima16} and $\text{F}_1$-ATPase angle~\cite{yasuda98}.
	
	When driving over a step potential~\cite{brown16}, there also is an optimal spring constant that maximizes time asymmetry relative to linear response. However, driving over steps exhibits increased time asymmetry for intermediate step heights, while driving over ratchets maximizes time asymmetry for large barrier heights. Sawtooth and step potentials are similar in that the crossing timescale for either a sufficiently high barrier or step may be longer than the protocol duration, leading to distinct trajectory distributions for forward and reverse protocols. However, moving the trap from the beginning to the end of the protocol for a step potential changes the equilibrium energy, while for a sawtooth potential the equilibrium energy remains unchanged. For a sufficiently large step, although the trap may be located on the high side of the step, most equilibrium probability will be on the low side of the step, allowing very similar trajectories for forward and reverse protocols; this is not possible for a sawtooth potential. This contrast between large steps and sawtooth barriers---regarding how equilibrium probability is located---causes time asymmetry to be maximal for intermediate step heights and large sawtooth barriers. 
	
	Previous work has focused on the effect of barrier shape on transition characteristics and molecular machine behavior~\cite{wagoner16,brown17,brown18}. Accordingly, we examined how an asymmetric sawtooth potential (with a barrier unevenly spaced between neighboring states) affects time asymmetry compared to a symmetric potential (barrier halfway between states). We found that symmetric ratchets maximize time asymmetry (Fig.~\ref{fig:asymmetry}). Symmetric ratchets maximize the distance between the sawtooth potential barrier maximum and the closest neighboring sawtooth potential minimum, decreasing the equilibrium probability to find the system on the opposite side of the barrier from the trap. This increases the chance of distinct forward and reverse trajectories and leads to higher time asymmetries.
	
	Our finding that time asymmetry (at a given dissipation) is maximized for symmetric sawtooth potentials (Fig.~\ref{fig:asymmetry}) contrasts with previous work on ratchet potentials for molecular machines, primarily with flashing ratchets~\cite{astumian94,julicher97,astumian98,parmeggiani99,lipowsky00,reimann02,suzuki03,linke05,kedem17,saakian18}.
	In particular, flashing ratchets model autonomous~\cite{machta15} behavior, and thus require spatially asymmetric sawteeth (or more generally a spatially asymmetric potential) to generate directed motion. In contrast, the deterministic protocols used in this work (i.e.\ the imposed trap translation schedules) externally provide direction, allowing the sawtooth to be symmetric while maintaining directed behavior. This difference between these two types of driving leads the sawtooth shape required for directed motion in flashing ratchets to be different from the sawtooth shape that optimizes the trade-off between time asymmetry and dissipation for deterministic protocols.
	
	Our results most directly apply to synthetic molecular machines, which are typically externally driven~\cite{erbas-cakmak15}. Although biomolecular machines are autonomous, some aspects of biomolecular machine operation may approximate an externally driven process: rapid changes, such as ATP binding or neck-linker docking for kinesin~\cite{muretta15b}, can occur spontaneously and drive a machine transition (distinct from the initial change) in a manner resembling a control protocol, and may yield similar design principles.

	Discrete-state models of molecular machines often describe the effect of a load (e.g.\ a constant resisting force) by slowing down forward transitions, with rate constant $k^+ \sim k^+_0 e^{-\delta\omega_{\text{load}}}$, or accelerating reverse transitions, $k^- \sim k^-_0 e^{-(1-\delta)\omega_{\text{load}}}$. Here $k^{\pm}$ are the transition rate constants, $k^{\pm}_0$ are the rate constants absent a load, and $\omega_{\text{load}}$ is the work done against the load over the transition~\cite{brown18}. The splitting factor $\delta$ quantitatively describes how the influence of the load is distributed to the forward and reverse transitions. $\delta$ can be thought of as reflecting the asymmetry of the transition-state position between the two metastable states~\cite{brown17} ($\delta=0$ indicates the transition state is close to the rear state, and $\delta=1$ indicates the transition state is close to the forward state), analogous to how the distance to the transition state affects unfolding rates~\cite{elms12}. Splitting factors have been estimated for a variety of biomolecular machines, and their values indicate that barriers between molecular machine states range from symmetric to very asymmetric~\cite{brown18}, suggesting that evolution could steer biomolecular machines towards symmetric potentials.
	
	\begin{acknowledgments}
		This work was supported by a Natural Sciences and Engineering Research Council of Canada (NSERC) Discovery Grant (D.A.S.), funds provided by the Faculty of Science, Simon Fraser University through the President's Research Start-up Grant (D.A.S.), a Tier-II Canada Research Chair (D.A.S.), an NSERC Undergraduate Student Research Award (A.Z.), and WestGrid (www.westgrid.ca) and Compute Canada Calcul Canada (www.computecanada.ca). The authors thank Steven J.~Large and Emma Lathouwers (SFU Physics) for useful discussions and feedback.
	\end{acknowledgments}

	\appendix
	\section{Nondimensionalization}
	\label{sec:nondimensionalization}
	
	The system energy $E'$ at position $x'$ in a quadratic trap with spring constant $k'$ centered at $x'=x_{\text{trap}}'$ with sawtooth barrier height $\HEIGHT'$ is
	\begin{equation}
	E' = \frac{1}{2}k'(x'-x_{\text{trap}}')^2 + \frac{\HEIGHT'x'}{\ell_1'} \ ,
	\end{equation}
	for $x'<\ell_1'$. We nondimensionalize all quantities using energy scale $H$ and length scale $\ell$:
	\begin{equation}
	\frac{E'}{H} = \frac{1}{2}\frac{k'\ell^2}{H}\frac{(x' - x_{\text{trap}}')^2}{\ell^2} + \frac{\HEIGHT'}{H}\frac{x'}{\ell}\frac{\ell}{\ell_1'} \ .
	\end{equation}
	We set $E = E'/H$, $k=k'\ell^2/H$, $x=x'/\ell$, $\HEIGHT=\HEIGHT'/H$, $\ell_1=\ell_1'/\ell$, and $\ell_2=\ell_2'/\ell$. We choose $H=k_{\text{B}}T$, so that all energies are in units of $k_{\text{B}}T$. $\ell$ is chosen so that $\ell_1 + \ell_2 = 1$, i.e.\ the nondimensionalized wavelength of the sawtooth is one. The same nondimensionalization applies for positions $x'>\ell_1'$. We also nondimensionalize all times by timescale $\tau$ chosen so that dimensionless diffusivity $D$ is unity.

	\section{Simulation details}
	\label{sec:simulation}
	
	The system moves on a one-dimensional spatial lattice with spacing $\Delta x_{\text{step}}=0.01$, with each sawtooth repeat $\ell=1$ in length. Each timestep $\Delta t_{\text{step}}$, the system randomly attempts to move either left or right, with moves accepted according to the Metropolis criterion~\cite{metropolis49}. $D=1$ is maintained by choosing the timestep, $\Delta t_{\text{step}} = (\Delta x_{\text{step}})^2/2D$. After each attempted system move, the trap takes a step of size $\Delta x_{\text{trap}} = u \, \Delta t_{\text{step}}$, where $u$ is the trap speed.

	\section{Instantaneous trap translation}
	\label{sec:instant}
	The system is on an infinitely repeating static sawtooth potential (sawtooth wavelength is one) and initially the trap is centered at $x_{\text{trap}}=0$. The energy at position $x$ is $E[x,x_{\text{trap}}(t)]=E_{\text{trap}}[x,x_{\text{trap}}(t)] + E_{\text{r}}(x)$. 
	If the trap moves instantaneously, the system does not move. When $x_{\text{trap}}$ is increased by one (the sawtooth wavelength), 
	\begin{equation}
	\Delta E_{\text{r}}=0 \ , \label{eq:dE}
	\end{equation}
	and $\Delta E_{\text{trap}}(x) = k(\tfrac{1}{2} - x)$. The mean work done by this instantaneous trap movement, the maximum that can be achieved for a monotonic protocol connecting the control-parameter endpoints, is
	\begin{equation}
	\label{eq:wmax}
	\langle W\rangle_{\text{max}}
	= \frac{1}{Z} \int_{-\infty}^{\infty}\Delta E_{\text{trap}}(x)e^{-E[x,x_{\text{trap}}=0]} {\mathrm d}x \ ,
	\end{equation}
	where $Z \equiv \int_{-\infty}^{\infty}e^{-E[x,x_{\text{trap}}=0]}{\mathrm d}x$. For a symmetric sawtooth potential, the ${\mathcal O}(x)$ term in Eq.~\eqref{eq:wmax} does not contribute to the integral, because the integral bounds are symmetric about $x=0$ and $kx \, e^{-E[x,x_{\text{trap}}=0]}$ is overall an odd function of $x$. Therefore $\langle W\rangle_{\text{max}}=\tfrac{1}{2}k$. The same result is found for instantaneous trap translation from $x=1$ to $x=0$ (reversing the protocol). Thus, using Eq.~\ref{eq:dissipation}, the maximum dissipation is $h_{\text{max}} = \langle W\rangle_{\text{max}} = \tfrac{1}{2}k$.
	
	%
	

\begin{thebibliography}{44}%
		\makeatletter
		\providecommand \@ifxundefined [1]{%
			\@ifx{#1\undefined}
		}%
		\providecommand \@ifnum [1]{%
			\ifnum #1\expandafter \@firstoftwo
			\else \expandafter \@secondoftwo
			\fi
		}%
		\providecommand \@ifx [1]{%
			\ifx #1\expandafter \@firstoftwo
			\else \expandafter \@secondoftwo
			\fi
		}%
		\providecommand \natexlab [1]{#1}%
		\providecommand \enquote  [1]{``#1''}%
		\providecommand \bibnamefont  [1]{#1}%
		\providecommand \bibfnamefont [1]{#1}%
		\providecommand \citenamefont [1]{#1}%
		\providecommand \href@noop [0]{\@secondoftwo}%
		\providecommand \href [0]{\begingroup \@sanitize@url \@href}%
		\providecommand \@href[1]{\@@startlink{#1}\@@href}%
		\providecommand \@@href[1]{\endgroup#1\@@endlink}%
		\providecommand \@sanitize@url [0]{\catcode `\\12\catcode `\$12\catcode
			`\&12\catcode `\#12\catcode `\^12\catcode `\_12\catcode `\%12\relax}%
		\providecommand \@@startlink[1]{}%
		\providecommand \@@endlink[0]{}%
		\providecommand \url  [0]{\begingroup\@sanitize@url \@url }%
		\providecommand \@url [1]{\endgroup\@href {#1}{\urlprefix }}%
		\providecommand \urlprefix  [0]{URL }%
		\providecommand \Eprint [0]{\href }%
		\providecommand \doibase [0]{http://dx.doi.org/}%
		\providecommand \selectlanguage [0]{\@gobble}%
		\providecommand \bibinfo  [0]{\@secondoftwo}%
		\providecommand \bibfield  [0]{\@secondoftwo}%
		\providecommand \translation [1]{[#1]}%
		\providecommand \BibitemOpen [0]{}%
		\providecommand \bibitemStop [0]{}%
		\providecommand \bibitemNoStop [0]{.\EOS\space}%
		\providecommand \EOS [0]{\spacefactor3000\relax}%
		\providecommand \BibitemShut  [1]{\csname bibitem#1\endcsname}%
		\let\auto@bib@innerbib\@empty
		\bibitem [{\citenamefont {Alberts}(1998)}]{alberts98}%
		\BibitemOpen
		\bibfield  {author} {\bibinfo {author} {\bibfnamefont {B.}~\bibnamefont
				{Alberts}},\ }\href@noop {} {\bibfield  {journal} {\bibinfo  {journal}
				{Cell}\ }\textbf {\bibinfo {volume} {92}},\ \bibinfo {pages} {291} (\bibinfo
			{year} {1998})}\BibitemShut {NoStop}%
		\bibitem [{\citenamefont {Zhang}\ \emph {et~al.}(2018)\citenamefont {Zhang},
			\citenamefont {Marcos},\ and\ \citenamefont {Leigh}}]{zhang18}%
		\BibitemOpen
		\bibfield  {author} {\bibinfo {author} {\bibfnamefont {L.}~\bibnamefont
				{Zhang}}, \bibinfo {author} {\bibfnamefont {V.}~\bibnamefont {Marcos}}, \
			and\ \bibinfo {author} {\bibfnamefont {D.~A.}\ \bibnamefont {Leigh}},\
		}\href@noop {} {\bibfield  {journal} {\bibinfo  {journal} {Proc.\ Natl.\
				Acad.\ Sci.\ USA}\ } (\bibinfo {year} {2018})}\BibitemShut {NoStop}%
	\bibitem [{\citenamefont {Seifert}(2012)}]{seifert12}%
	\BibitemOpen
	\bibfield  {author} {\bibinfo {author} {\bibfnamefont {U.}~\bibnamefont
			{Seifert}},\ }\href@noop {} {\bibfield  {journal} {\bibinfo  {journal} {Rep.\
				Prog.\ Phys.}\ }\textbf {\bibinfo {volume} {75}},\ \bibinfo {pages} {126001}
		(\bibinfo {year} {2012})}\BibitemShut {NoStop}%
	\bibitem [{\citenamefont {Brown}\ and\ \citenamefont
		{Sivak}(2017{\natexlab{a}})}]{brown17b}%
	\BibitemOpen
	\bibfield  {author} {\bibinfo {author} {\bibfnamefont {A.~I.}\ \bibnamefont
			{Brown}}\ and\ \bibinfo {author} {\bibfnamefont {D.~A.}\ \bibnamefont
			{Sivak}},\ }\href@noop {} {\bibfield  {journal} {\bibinfo  {journal} {Physics
				in Canada}\ }\textbf {\bibinfo {volume} {73}},\ \bibinfo {pages} {61}
		(\bibinfo {year} {2017}{\natexlab{a}})}\BibitemShut {NoStop}%
	\bibitem [{\citenamefont {den Broeck}\ \emph {et~al.}(2012)\citenamefont {den
			Broeck}, \citenamefont {Kumar},\ and\ \citenamefont
		{Lindenberg}}]{vandenbroeck12}%
	\BibitemOpen
	\bibfield  {author} {\bibinfo {author} {\bibfnamefont {C.~V.}\ \bibnamefont
			{den Broeck}}, \bibinfo {author} {\bibfnamefont {N.}~\bibnamefont {Kumar}}, \
		and\ \bibinfo {author} {\bibfnamefont {K.}~\bibnamefont {Lindenberg}},\
	}\href@noop {} {\bibfield  {journal} {\bibinfo  {journal} {Phys.\ Rev.\
			Lett.}\ }\textbf {\bibinfo {volume} {108}},\ \bibinfo {pages} {210602}
	(\bibinfo {year} {2012})}\BibitemShut {NoStop}%
\bibitem [{\citenamefont {Schmiedl}\ and\ \citenamefont
	{Seifert}(2008)}]{schmiedl08}%
\BibitemOpen
\bibfield  {author} {\bibinfo {author} {\bibfnamefont {T.}~\bibnamefont
		{Schmiedl}}\ and\ \bibinfo {author} {\bibfnamefont {U.}~\bibnamefont
		{Seifert}},\ }\href@noop {} {\bibfield  {journal} {\bibinfo  {journal}
		{Europhys.\ Lett.}\ }\textbf {\bibinfo {volume} {83}},\ \bibinfo {pages}
	{30005} (\bibinfo {year} {2008})}\BibitemShut {NoStop}%
\bibitem [{\citenamefont {Wagoner}\ and\ \citenamefont
	{Dill}(2016)}]{wagoner16}%
\BibitemOpen
\bibfield  {author} {\bibinfo {author} {\bibfnamefont {J.~A.}\ \bibnamefont
		{Wagoner}}\ and\ \bibinfo {author} {\bibfnamefont {K.~A.}\ \bibnamefont
		{Dill}},\ }\href@noop {} {\bibfield  {journal} {\bibinfo  {journal} {J.\
			Phys.\ Chem.\ B}\ }\textbf {\bibinfo {volume} {120}},\ \bibinfo {pages}
	{6327} (\bibinfo {year} {2016})}\BibitemShut {NoStop}%
\bibitem [{\citenamefont {Brown}\ and\ \citenamefont
	{Sivak}(2017{\natexlab{b}})}]{brown17}%
\BibitemOpen
\bibfield  {author} {\bibinfo {author} {\bibfnamefont {A.~I.}\ \bibnamefont
		{Brown}}\ and\ \bibinfo {author} {\bibfnamefont {D.~A.}\ \bibnamefont
		{Sivak}},\ }\href@noop {} {\bibfield  {journal} {\bibinfo  {journal} {Proc.\
			Natl.\ Acad.\ Sci.\ USA}\ }\textbf {\bibinfo {volume} {114}},\ \bibinfo
	{pages} {11057} (\bibinfo {year} {2017}{\natexlab{b}})}\BibitemShut {NoStop}%
\bibitem [{\citenamefont {Brown}\ and\ \citenamefont {Sivak}(2018)}]{brown18}%
\BibitemOpen
\bibfield  {author} {\bibinfo {author} {\bibfnamefont {A.~I.}\ \bibnamefont
		{Brown}}\ and\ \bibinfo {author} {\bibfnamefont {D.~A.}\ \bibnamefont
		{Sivak}},\ }\href@noop {} {\bibfield  {journal} {\bibinfo  {journal} {J.\
			Phys.\ Chem.\ B}\ }\textbf {\bibinfo {volume} {122}},\ \bibinfo {pages}
	{1387} (\bibinfo {year} {2018})}\BibitemShut {NoStop}%
\bibitem [{\citenamefont {Barato}\ and\ \citenamefont
	{Seifert}(2015)}]{barato15}%
\BibitemOpen
\bibfield  {author} {\bibinfo {author} {\bibfnamefont {A.~C.}\ \bibnamefont
		{Barato}}\ and\ \bibinfo {author} {\bibfnamefont {U.}~\bibnamefont
		{Seifert}},\ }\href@noop {} {\bibfield  {journal} {\bibinfo  {journal}
		{Phys.\ Rev.\ Lett.}\ }\textbf {\bibinfo {volume} {114}},\ \bibinfo {pages}
	{158101} (\bibinfo {year} {2015})}\BibitemShut {NoStop}%
\bibitem [{\citenamefont {Machta}(2015)}]{machta15}%
\BibitemOpen
\bibfield  {author} {\bibinfo {author} {\bibfnamefont {B.~B.}\ \bibnamefont
		{Machta}},\ }\href@noop {} {\bibfield  {journal} {\bibinfo  {journal} {Phys.\
			Rev.\ Lett.}\ }\textbf {\bibinfo {volume} {115}},\ \bibinfo {pages} {260603}
	(\bibinfo {year} {2015})}\BibitemShut {NoStop}%
\bibitem [{\citenamefont {Ariga}\ \emph {et~al.}(2018)\citenamefont {Ariga},
	\citenamefont {Tomishige},\ and\ \citenamefont {Mizuno}}]{ariga18}%
\BibitemOpen
\bibfield  {author} {\bibinfo {author} {\bibfnamefont {T.}~\bibnamefont
		{Ariga}}, \bibinfo {author} {\bibfnamefont {M.}~\bibnamefont {Tomishige}}, \
	and\ \bibinfo {author} {\bibfnamefont {D.}~\bibnamefont {Mizuno}},\
}\href@noop {} {\bibfield  {journal} {\bibinfo  {journal} {Phys.\ Rev.\
		Lett.}\ }\textbf {\bibinfo {volume} {121}},\ \bibinfo {pages} {218101}
(\bibinfo {year} {2018})}\BibitemShut {NoStop}%
\bibitem [{\citenamefont {Feng}\ and\ \citenamefont {Crooks}(2008)}]{feng08}%
\BibitemOpen
\bibfield  {author} {\bibinfo {author} {\bibfnamefont {E.~H.}\ \bibnamefont
		{Feng}}\ and\ \bibinfo {author} {\bibfnamefont {G.~E.}\ \bibnamefont
		{Crooks}},\ }\href@noop {} {\bibfield  {journal} {\bibinfo  {journal} {Phys.\
			Rev.\ Lett.}\ }\textbf {\bibinfo {volume} {101}},\ \bibinfo {pages} {090602}
	(\bibinfo {year} {2008})}\BibitemShut {NoStop}%
\bibitem [{\citenamefont {Procacci}\ and\ \citenamefont
	{Marsili}(2010)}]{procacci10}%
\BibitemOpen
\bibfield  {author} {\bibinfo {author} {\bibfnamefont {P.}~\bibnamefont
		{Procacci}}\ and\ \bibinfo {author} {\bibfnamefont {S.}~\bibnamefont
		{Marsili}},\ }\href@noop {} {\bibfield  {journal} {\bibinfo  {journal}
		{Chem.\ Phys.}\ }\textbf {\bibinfo {volume} {375}},\ \bibinfo {pages} {8}
	(\bibinfo {year} {2010})}\BibitemShut {NoStop}%
\bibitem [{\citenamefont {Marconi}(2008)}]{marconi08}%
\BibitemOpen
\bibfield  {author} {\bibinfo {author} {\bibfnamefont {U.~M.~B.}\
		\bibnamefont {Marconi}},\ }\href@noop {} {\bibfield  {journal} {\bibinfo
		{journal} {Phys.\ Rep.}\ }\textbf {\bibinfo {volume} {461}},\ \bibinfo
	{pages} {111} (\bibinfo {year} {2008})}\BibitemShut {NoStop}%
\bibitem [{\citenamefont {Brown}\ and\ \citenamefont {Sivak}(2016)}]{brown16}%
\BibitemOpen
\bibfield  {author} {\bibinfo {author} {\bibfnamefont {A.~I.}\ \bibnamefont
		{Brown}}\ and\ \bibinfo {author} {\bibfnamefont {D.~A.}\ \bibnamefont
		{Sivak}},\ }\href@noop {} {\bibfield  {journal} {\bibinfo  {journal} {Phys.\
			Rev.\ E}\ }\textbf {\bibinfo {volume} {94}},\ \bibinfo {pages} {032137}
	(\bibinfo {year} {2016})}\BibitemShut {NoStop}%
\bibitem [{\citenamefont {Astumian}\ and\ \citenamefont
	{Bier}(1994)}]{astumian94}%
\BibitemOpen
\bibfield  {author} {\bibinfo {author} {\bibfnamefont {R.~D.}\ \bibnamefont
		{Astumian}}\ and\ \bibinfo {author} {\bibfnamefont {M.}~\bibnamefont
		{Bier}},\ }\href@noop {} {\bibfield  {journal} {\bibinfo  {journal} {Phys.\
			Rev.\ Lett.}\ }\textbf {\bibinfo {volume} {72}},\ \bibinfo {pages} {1766}
	(\bibinfo {year} {1994})}\BibitemShut {NoStop}%
\bibitem [{\citenamefont {Julicher}\ \emph {et~al.}(1997)\citenamefont
	{Julicher}, \citenamefont {Ajdari},\ and\ \citenamefont
	{Prost}}]{julicher97}%
\BibitemOpen
\bibfield  {author} {\bibinfo {author} {\bibfnamefont {F.}~\bibnamefont
		{Julicher}}, \bibinfo {author} {\bibfnamefont {A.}~\bibnamefont {Ajdari}}, \
	and\ \bibinfo {author} {\bibfnamefont {J.}~\bibnamefont {Prost}},\
}\href@noop {} {\bibfield  {journal} {\bibinfo  {journal} {Rev.\ Mod.\
		Phys.}\ }\textbf {\bibinfo {volume} {69}},\ \bibinfo {pages} {1269} (\bibinfo
{year} {1997})}\BibitemShut {NoStop}%
\bibitem [{\citenamefont {Astumian}\ and\ \citenamefont
	{Derenyi}(1998)}]{astumian98}%
\BibitemOpen
\bibfield  {author} {\bibinfo {author} {\bibfnamefont {R.~D.}\ \bibnamefont
		{Astumian}}\ and\ \bibinfo {author} {\bibfnamefont {I.}~\bibnamefont
		{Derenyi}},\ }\href@noop {} {\bibfield  {journal} {\bibinfo  {journal} {Eur.\
			Biophys.\ J.}\ }\textbf {\bibinfo {volume} {27}},\ \bibinfo {pages} {474}
	(\bibinfo {year} {1998})}\BibitemShut {NoStop}%
\bibitem [{\citenamefont {Parmeggiani}\ \emph {et~al.}(1999)\citenamefont
	{Parmeggiani}, \citenamefont {Julicher}, \citenamefont {Ajdari},\ and\
	\citenamefont {Prost}}]{parmeggiani99}%
\BibitemOpen
\bibfield  {author} {\bibinfo {author} {\bibfnamefont {A.}~\bibnamefont
		{Parmeggiani}}, \bibinfo {author} {\bibfnamefont {F.}~\bibnamefont
		{Julicher}}, \bibinfo {author} {\bibfnamefont {A.}~\bibnamefont {Ajdari}}, \
	and\ \bibinfo {author} {\bibfnamefont {J.}~\bibnamefont {Prost}},\
}\href@noop {} {\bibfield  {journal} {\bibinfo  {journal} {Phys.\ Rev.\ E}\
}\textbf {\bibinfo {volume} {60}},\ \bibinfo {pages} {2127} (\bibinfo {year}
{1999})}\BibitemShut {NoStop}%
\bibitem [{\citenamefont {Lipowsky}\ and\ \citenamefont
	{Harms}(2000)}]{lipowsky00}%
\BibitemOpen
\bibfield  {author} {\bibinfo {author} {\bibfnamefont {R.}~\bibnamefont
		{Lipowsky}}\ and\ \bibinfo {author} {\bibfnamefont {T.}~\bibnamefont
		{Harms}},\ }\href@noop {} {\bibfield  {journal} {\bibinfo  {journal} {Eur.\
			Biophys.\ J.}\ }\textbf {\bibinfo {volume} {29}},\ \bibinfo {pages} {542}
	(\bibinfo {year} {2000})}\BibitemShut {NoStop}%
\bibitem [{\citenamefont {Reimann}(2002)}]{reimann02}%
\BibitemOpen
\bibfield  {author} {\bibinfo {author} {\bibfnamefont {P.}~\bibnamefont
		{Reimann}},\ }\href@noop {} {\bibfield  {journal} {\bibinfo  {journal}
		{Phys.\ Rep.}\ }\textbf {\bibinfo {volume} {361}},\ \bibinfo {pages} {57}
	(\bibinfo {year} {2002})}\BibitemShut {NoStop}%
\bibitem [{\citenamefont {Suzuki}\ and\ \citenamefont
	{Munakata}(2003)}]{suzuki03}%
\BibitemOpen
\bibfield  {author} {\bibinfo {author} {\bibfnamefont {D.}~\bibnamefont
		{Suzuki}}\ and\ \bibinfo {author} {\bibfnamefont {T.}~\bibnamefont
		{Munakata}},\ }\href@noop {} {\bibfield  {journal} {\bibinfo  {journal}
		{Phys.\ Rev.\ E}\ }\textbf {\bibinfo {volume} {68}},\ \bibinfo {pages}
	{021906} (\bibinfo {year} {2003})}\BibitemShut {NoStop}%
\bibitem [{\citenamefont {Gabrys}\ \emph {et~al.}(2004)\citenamefont {Gabrys},
	\citenamefont {Pesz},\ and\ \citenamefont {Bartkiewicz}}]{gabrys04}%
\BibitemOpen
\bibfield  {author} {\bibinfo {author} {\bibfnamefont {B.~J.}\ \bibnamefont
		{Gabrys}}, \bibinfo {author} {\bibfnamefont {K.}~\bibnamefont {Pesz}}, \ and\
	\bibinfo {author} {\bibfnamefont {S.~J.}\ \bibnamefont {Bartkiewicz}},\
}\href@noop {} {\bibfield  {journal} {\bibinfo  {journal} {Physica A}\
}\textbf {\bibinfo {volume} {336}},\ \bibinfo {pages} {112} (\bibinfo {year}
{2004})}\BibitemShut {NoStop}%
\bibitem [{\citenamefont {Linke}\ \emph {et~al.}(2005)\citenamefont {Linke},
	\citenamefont {Downton},\ and\ \citenamefont {Zuckermann}}]{linke05}%
\BibitemOpen
\bibfield  {author} {\bibinfo {author} {\bibfnamefont {H.}~\bibnamefont
		{Linke}}, \bibinfo {author} {\bibfnamefont {M.~T.}\ \bibnamefont {Downton}},
	\ and\ \bibinfo {author} {\bibfnamefont {M.~J.}\ \bibnamefont {Zuckermann}},\
}\href@noop {} {\bibfield  {journal} {\bibinfo  {journal} {Chaos}\ }\textbf
{\bibinfo {volume} {15}},\ \bibinfo {pages} {026111} (\bibinfo {year}
{2005})}\BibitemShut {NoStop}%
\bibitem [{\citenamefont {Lacoste}\ and\ \citenamefont
	{Mallick}(2009)}]{lacoste09}%
\BibitemOpen
\bibfield  {author} {\bibinfo {author} {\bibfnamefont {D.}~\bibnamefont
		{Lacoste}}\ and\ \bibinfo {author} {\bibfnamefont {K.}~\bibnamefont
		{Mallick}},\ }\href@noop {} {\bibfield  {journal} {\bibinfo  {journal}
		{Phys.\ Rev.\ Lett.}\ }\textbf {\bibinfo {volume} {80}},\ \bibinfo {pages}
	{021923} (\bibinfo {year} {2009})}\BibitemShut {NoStop}%
\bibitem [{\citenamefont {Levien}\ and\ \citenamefont
	{Bressloff}(2015)}]{levien15}%
\BibitemOpen
\bibfield  {author} {\bibinfo {author} {\bibfnamefont {E.}~\bibnamefont
		{Levien}}\ and\ \bibinfo {author} {\bibfnamefont {P.~C.}\ \bibnamefont
		{Bressloff}},\ }\href@noop {} {\bibfield  {journal} {\bibinfo  {journal}
		{Phys.\ Rev.\ E}\ }\textbf {\bibinfo {volume} {92}},\ \bibinfo {pages}
	{042129} (\bibinfo {year} {2015})}\BibitemShut {NoStop}%
\bibitem [{\citenamefont {Kedem}\ \emph {et~al.}(2017)\citenamefont {Kedem},
	\citenamefont {Lau},\ and\ \citenamefont {Weiss}}]{kedem17}%
\BibitemOpen
\bibfield  {author} {\bibinfo {author} {\bibfnamefont {O.}~\bibnamefont
		{Kedem}}, \bibinfo {author} {\bibfnamefont {B.}~\bibnamefont {Lau}}, \ and\
	\bibinfo {author} {\bibfnamefont {E.~A.}\ \bibnamefont {Weiss}},\ }\href@noop
{} {\bibfield  {journal} {\bibinfo  {journal} {ACS Nano}\ }\textbf {\bibinfo
		{volume} {11}},\ \bibinfo {pages} {7148} (\bibinfo {year}
	{2017})}\BibitemShut {NoStop}%
\bibitem [{\citenamefont {Saakian}\ and\ \citenamefont
	{Klumper}(2018)}]{saakian18}%
\BibitemOpen
\bibfield  {author} {\bibinfo {author} {\bibfnamefont {D.~B.}\ \bibnamefont
		{Saakian}}\ and\ \bibinfo {author} {\bibfnamefont {A.}~\bibnamefont
		{Klumper}},\ }\href@noop {} {\bibfield  {journal} {\bibinfo  {journal}
		{Phys.\ Rev.\ E}\ }\textbf {\bibinfo {volume} {97}},\ \bibinfo {pages}
	{012153} (\bibinfo {year} {2018})}\BibitemShut {NoStop}%
\bibitem [{\citenamefont {Crooks}\ and\ \citenamefont
	{Sivak}(2011)}]{crooks11}%
\BibitemOpen
\bibfield  {author} {\bibinfo {author} {\bibfnamefont {G.~E.}\ \bibnamefont
		{Crooks}}\ and\ \bibinfo {author} {\bibfnamefont {D.~A.}\ \bibnamefont
		{Sivak}},\ }\href@noop {} {\bibfield  {journal} {\bibinfo  {journal} {J.\
			Stat.\ Mech.: Theory Exp.}\ ,\ \bibinfo {pages} {P06003}} (\bibinfo {year}
	{2011})}\BibitemShut {NoStop}%
\bibitem [{\citenamefont {Crooks}(1999)}]{crooks99}%
\BibitemOpen
\bibfield  {author} {\bibinfo {author} {\bibfnamefont {G.~E.}\ \bibnamefont
		{Crooks}},\ }\href@noop {} {\bibfield  {journal} {\bibinfo  {journal} {Phys.\
			Rev.\ E}\ }\textbf {\bibinfo {volume} {60}},\ \bibinfo {pages} {2721}
	(\bibinfo {year} {1999})}\BibitemShut {NoStop}%
\bibitem [{\citenamefont {Taneja}(2005)}]{taneja05}%
\BibitemOpen
\bibfield  {author} {\bibinfo {author} {\bibfnamefont {I.~J.}\ \bibnamefont
		{Taneja}},\ }\href@noop {} {\bibfield  {journal} {\bibinfo  {journal} {Aust.\
			J.\ Math.\ Anal.}\ }\textbf {\bibinfo {volume} {2}},\ \bibinfo {pages} {1}
	(\bibinfo {year} {2005})}\BibitemShut {NoStop}%
\bibitem [{\citenamefont {Chandler}(1987)}]{chandler87}%
\BibitemOpen
\bibfield  {author} {\bibinfo {author} {\bibfnamefont {D.}~\bibnamefont
		{Chandler}},\ }\href@noop {} {\emph {\bibinfo {title} {{Introduction to
				modern statistical mechanics}}}}\ (\bibinfo  {publisher} {Oxford University
	Press},\ \bibinfo {year} {1987})\BibitemShut {NoStop}%
\bibitem [{\citenamefont {Speck}\ and\ \citenamefont
	{Seifert}(2004)}]{speck04}%
\BibitemOpen
\bibfield  {author} {\bibinfo {author} {\bibfnamefont {T.}~\bibnamefont
		{Speck}}\ and\ \bibinfo {author} {\bibfnamefont {U.}~\bibnamefont
		{Seifert}},\ }\href@noop {} {\bibfield  {journal} {\bibinfo  {journal}
		{Phys.\ Rev.\ E}\ }\textbf {\bibinfo {volume} {70}},\ \bibinfo {pages}
	{066112} (\bibinfo {year} {2004})}\BibitemShut {NoStop}%
\bibitem [{\citenamefont {Grabert}(1988)}]{grabert88}%
\BibitemOpen
\bibfield  {author} {\bibinfo {author} {\bibfnamefont {H.}~\bibnamefont
		{Grabert}},\ }\href@noop {} {\bibfield  {journal} {\bibinfo  {journal}
		{Phys.\ Rev.\ Lett.}\ }\textbf {\bibinfo {volume} {61}},\ \bibinfo {pages}
	{1683} (\bibinfo {year} {1988})}\BibitemShut {NoStop}%
\bibitem [{\citenamefont {Chang}\ \emph {et~al.}(2013)\citenamefont {Chang},
	\citenamefont {Nitta}, \citenamefont {Inoue},\ and\ \citenamefont
	{Hirokawa}}]{chang13}%
\BibitemOpen
\bibfield  {author} {\bibinfo {author} {\bibfnamefont {Q.}~\bibnamefont
		{Chang}}, \bibinfo {author} {\bibfnamefont {R.}~\bibnamefont {Nitta}},
	\bibinfo {author} {\bibfnamefont {S.}~\bibnamefont {Inoue}}, \ and\ \bibinfo
	{author} {\bibfnamefont {N.}~\bibnamefont {Hirokawa}},\ }\href@noop {}
{\bibfield  {journal} {\bibinfo  {journal} {J.\ Mol.\ Biol.}\ }\textbf
	{\bibinfo {volume} {425}},\ \bibinfo {pages} {1869} (\bibinfo {year}
	{2013})}\BibitemShut {NoStop}%
\bibitem [{\citenamefont {Isojima}\ \emph {et~al.}(2016)\citenamefont
	{Isojima}, \citenamefont {Iino}, \citenamefont {Niitani}, \citenamefont
	{Noji},\ and\ \citenamefont {Tomishige}}]{isojima16}%
\BibitemOpen
\bibfield  {author} {\bibinfo {author} {\bibfnamefont {H.}~\bibnamefont
		{Isojima}}, \bibinfo {author} {\bibfnamefont {R.}~\bibnamefont {Iino}},
	\bibinfo {author} {\bibfnamefont {Y.}~\bibnamefont {Niitani}}, \bibinfo
	{author} {\bibfnamefont {H.}~\bibnamefont {Noji}}, \ and\ \bibinfo {author}
	{\bibfnamefont {M.}~\bibnamefont {Tomishige}},\ }\href@noop {} {\bibfield
	{journal} {\bibinfo  {journal} {Nat.\ Chem.\ Biol.}\ }\textbf {\bibinfo
		{volume} {12}},\ \bibinfo {pages} {290} (\bibinfo {year} {2016})}\BibitemShut
{NoStop}%
\bibitem [{\citenamefont {Hackney}(1988)}]{hackney88}%
\BibitemOpen
\bibfield  {author} {\bibinfo {author} {\bibfnamefont {D.~D.}\ \bibnamefont
		{Hackney}},\ }\href@noop {} {\bibfield  {journal} {\bibinfo  {journal}
		{Proc.\ Natl.\ Acad.\ Sci.\ USA}\ }\textbf {\bibinfo {volume} {85}},\
	\bibinfo {pages} {6314} (\bibinfo {year} {1988})}\BibitemShut {NoStop}%
\bibitem [{\citenamefont {Nogales}\ and\ \citenamefont
	{Zhang}(2016)}]{nogales16}%
\BibitemOpen
\bibfield  {author} {\bibinfo {author} {\bibfnamefont {E.}~\bibnamefont
		{Nogales}}\ and\ \bibinfo {author} {\bibfnamefont {R.}~\bibnamefont
		{Zhang}},\ }\href@noop {} {\bibfield  {journal} {\bibinfo  {journal} {Curr.\
			Opin.\ Struct.\ Biol.}\ }\textbf {\bibinfo {volume} {37}},\ \bibinfo {pages}
	{90} (\bibinfo {year} {2016})}\BibitemShut {NoStop}%
\bibitem [{\citenamefont {Yasuda}\ \emph {et~al.}(1998)\citenamefont {Yasuda},
	\citenamefont {Noji}, \citenamefont {Kinosita},\ and\ \citenamefont
	{Yoshida}}]{yasuda98}%
\BibitemOpen
\bibfield  {author} {\bibinfo {author} {\bibfnamefont {R.}~\bibnamefont
		{Yasuda}}, \bibinfo {author} {\bibfnamefont {H.}~\bibnamefont {Noji}},
	\bibinfo {author} {\bibfnamefont {K.}~\bibnamefont {Kinosita}}, \ and\
	\bibinfo {author} {\bibfnamefont {M.}~\bibnamefont {Yoshida}},\ }\href@noop
{} {\bibfield  {journal} {\bibinfo  {journal} {Cell}\ ,\ \bibinfo {pages}
		{1117}} (\bibinfo {year} {1998})}\BibitemShut {NoStop}%
\bibitem [{\citenamefont {Erbas-Cakmak}\ \emph {et~al.}(2015)\citenamefont
	{Erbas-Cakmak}, \citenamefont {Leigh}, \citenamefont {McTernan},\ and\
	\citenamefont {Nussbaumer}}]{erbas-cakmak15}%
\BibitemOpen
\bibfield  {author} {\bibinfo {author} {\bibfnamefont {S.}~\bibnamefont
		{Erbas-Cakmak}}, \bibinfo {author} {\bibfnamefont {D.~A.}\ \bibnamefont
		{Leigh}}, \bibinfo {author} {\bibfnamefont {C.~T.}\ \bibnamefont {McTernan}},
	\ and\ \bibinfo {author} {\bibfnamefont {A.~L.}\ \bibnamefont {Nussbaumer}},\
}\href@noop {} {\bibfield  {journal} {\bibinfo  {journal} {Chem.\ Rev.}\
}\textbf {\bibinfo {volume} {115}},\ \bibinfo {pages} {10081} (\bibinfo
{year} {2015})}\BibitemShut {NoStop}%
\bibitem [{\citenamefont {Muretta}\ \emph {et~al.}(2015)\citenamefont
	{Muretta}, \citenamefont {Jun}, \citenamefont {Gross}, \citenamefont {Major},
	\citenamefont {Thomas},\ and\ \citenamefont {Rosenfeld}}]{muretta15b}%
\BibitemOpen
\bibfield  {author} {\bibinfo {author} {\bibfnamefont {J.~M.}\ \bibnamefont
		{Muretta}}, \bibinfo {author} {\bibfnamefont {Y.}~\bibnamefont {Jun}},
	\bibinfo {author} {\bibfnamefont {S.~P.}\ \bibnamefont {Gross}}, \bibinfo
	{author} {\bibfnamefont {J.}~\bibnamefont {Major}}, \bibinfo {author}
	{\bibfnamefont {D.~D.}\ \bibnamefont {Thomas}}, \ and\ \bibinfo {author}
	{\bibfnamefont {S.~S.}\ \bibnamefont {Rosenfeld}},\ }\href@noop {} {\bibfield
	{journal} {\bibinfo  {journal} {Proc.\ Natl.\ Acad.\ Sci.\ USA}\ }\textbf
	{\bibinfo {volume} {112}},\ \bibinfo {pages} {E6606} (\bibinfo {year}
	{2015})}\BibitemShut {NoStop}%
\bibitem [{\citenamefont {Elms}\ \emph {et~al.}(2012)\citenamefont {Elms},
	\citenamefont {Chodera}, \citenamefont {Bustamante},\ and\ \citenamefont
	{Marqusee}}]{elms12}%
\BibitemOpen
\bibfield  {author} {\bibinfo {author} {\bibfnamefont {P.~J.}\ \bibnamefont
		{Elms}}, \bibinfo {author} {\bibfnamefont {J.~D.}\ \bibnamefont {Chodera}},
	\bibinfo {author} {\bibfnamefont {C.}~\bibnamefont {Bustamante}}, \ and\
	\bibinfo {author} {\bibfnamefont {S.}~\bibnamefont {Marqusee}},\ }\href@noop
{} {\bibfield  {journal} {\bibinfo  {journal} {Proc.\ Natl.\ Acad.\ Sci.\
			USA}\ }\textbf {\bibinfo {volume} {109}},\ \bibinfo {pages} {3796} (\bibinfo
	{year} {2012})}\BibitemShut {NoStop}%
\bibitem [{\citenamefont {Metropolis}\ and\ \citenamefont
	{Ulam}(1949)}]{metropolis49}%
\BibitemOpen
\bibfield  {author} {\bibinfo {author} {\bibfnamefont {N.}~\bibnamefont
		{Metropolis}}\ and\ \bibinfo {author} {\bibfnamefont {S.}~\bibnamefont
		{Ulam}},\ }\href@noop {} {\bibfield  {journal} {\bibinfo  {journal} {J.\ Am.\
			Stat.\ Assoc.}\ }\textbf {\bibinfo {volume} {44}},\ \bibinfo {pages} {335}
	(\bibinfo {year} {1949})}\BibitemShut {NoStop}%
\end{thebibliography}

\end{document}